# The PLATO Solar-like Light-curve Simulator

## A tool to generate realistic stellar light-curves with instrumental effects representative of the PLATO mission

R. Samadi[1], A. Deru[1], D. Reese[1], V. Marchiori[1,2], E. Grolleau[1], J.J. Green[1], M. Pertenais[3], Y. Lebreton[1,4], S. Deheuvels[5], B. Mosser[1], K. Belkacem[1] , A. Börner[3], A. M. S. Smith[6]

[1] LESIA, Observatoire de Paris, Universit PSL, CNRS, Sorbonne Universit, Univ. Paris Diderot, Sorbonne Paris Cit, 5 place Jules Janssen, 92195 Meudon, France

[2] Escola Politécnica – Departamento de Engenharia de Telecomunicações e Controle, Universidade de São Paulo, Av. Prof. Luciano Gualberto, 05508-010 São Paulo, Brazil.

[3] Institute of Optical Sensors Systems, German Aerospace Center (DLR), Rutherfordstrasse 2, 12489 Berlin, Germany

[4] Univ Rennes, CNRS, IPR (Institut de Physique de Rennes) - UMR 6251, F-35000 Rennes, France

[5] IRAP, Universit de Toulouse, CNRS, CNES, UPS, Toulouse, France

[6] Institute of Planetary Research, German Aerospace Center, Rutherfordstrasse 2, 12489 Berlin, Germany

March 26, 2019

**ABSTRACT**

*Context.* ESA's PLATO space mission, to be launched by the end of 2026, aims to detect and characterise earth-like planets in their habitable zone using asteroseismology and the analysis of the transit events. The preparation of science objectives will require the implementation of hare-and-hound exercises relying on the massive generation of representative simulated light-curves.

*Aims.* We developed a light-curve simulator named the PLATO Solar-like Light-curve Simulator (PSLS) in order to generate light-curves representative of typical PLATO targets, that is showing simultaneously solar-like oscillations, stellar granulation, and magnetic activity. At the same time, PSLS also aims at mimicking in a realistic way the random noise and the systematic errors representative of the PLATO multi-telescope concept.

*Methods.* To quantify the instrumental systematic errors, we performed a series of simulations at pixel level that include various relevant sources of perturbations expected for PLATO. From the simulated pixels, we extract the photometry as planned on-board and also simulate the quasi-periodic updates of the aperture masks during the observations. The simulated light-curves are then corrected for instrumental effects using the instrument point spread functions reconstructed on the basis of a microscanning technique that will be operated during the in-flight calibration phases of the mission. These corrected and simulated light-curves are then fitted by a parametric model, which we incorporated in PSLS. Simulation of the oscillations and granulation signals rely on current state-of-the-art stellar seismology.

*Results.* We show that the instrumental systematic errors dominate the signal only at frequencies below ∼ 20 μHz. The systematic errors level is found to mainly depend on stellar magnitude and on the detector charge transfer inefficiency. To illustrate how realistic our simulator is, we compared its predictions with observations made by *Kepler* on three typical targets and found a good qualitative agreement with the observations.

*Conclusions.* PSLS reproduces the main properties of expected PLATO light-curves. Its speed of execution and its inclusion of relevant stellar signals as well as sources of noises representative of the PLATO cameras make it an indispensable tool for the scientific preparation of the PLATO mission.

## 1. Introduction

ESA's PLATO[1] space mission is expected to be launched by the end of 2026 with the goal of detecting and characterising earth-like planets in the habitable zone of dwarf and sub-giant stars of spectral types F to K (Rauer et al. 2014). The age and mass of planet-hosting stars will be determined by applying stellar seismic techniques to their solar-like oscillations (see e.g. Gizon et al. 2013; Van Eylen et al. 2014, 2018; Huber et al. 2019). The determination of these stellar parameters is a complex procedure since it relies on both the precise seismic analysis of the individual mode frequencies and the use of sophisticated stellar modelling techniques (see e.g. Lebreton et al. 2014a,b). To develop and test such complex procedures, realistic simulated light-curves are needed. These simulated light-curves are, for instance, typically used to conduct hare-and-hounds exercises[2] involving various teams in charge of the seismic analysis and stellar modelling (see e.g. Reese et al. 2016, and references therein). They are also used to conduct massive Monte Carlo simulations that enable one to assess the performances of seismic analysis pipelines (e.g. de Assis Peralta et al. 2018, and reference therein). The simulated light-curves must be sufficiently real-istic to accurately account for the properties of the modes but also for the other sources of stellar noise such as the granulation noise and the instrumental random noise that – to a large extent – limit the precision of the age and mass determination. Similar hare-and-hounds exercises are also planned to be carried out to test the efficiency of planet detection and the accuracy of the derived transit parameters. Since planetary transits are expected to last several hours, their analysis is quite sensitive to the noises occurring at low frequencies (typically below a few ten of $\mu$Hz). Finally, simulated light-curves are also used to prepare the analysis of the PLATO light-curves for a variety of other scientific objectives that are also relevant at low frequencies. We can, for instance, mention the characterisation of stellar granulation, the detection and characterisation of rotational modulations, among others. Accordingly, it is necessary to simulate in a realistic way the different sources of noise that dominate the signal at low frequencies. Among them, we have predominantly the stellar activity signal, but systematic instrumental errors may also intervene.

---

[2] Hare-and-hounds exercises typically involve several teams: one team produces a set of artificial observations while the other teams try to infer the physical model/properties behind these observations.

The CoRoT (Baglin et al. 2006b,a) and *Kepler* (Borucki et al. 2010) space missions, allowed us to carry out seismic studies of several thousands of pulsating red-giant stars (De Ridder et al. 2009; Kallinger et al. 2010; Stello et al. 2013) thus enabling important progress in our understanding of stellar interiors (see e.g. the reviews by Mosser & Miglio 2016; Hekker & Christensen-Dalsgaard 2017). These observations opened up the path to what we now call ensemble asteroseismology (see e.g. Huber et al. 2011; Belkacem et al. 2013; Miglio et al. 2015) with various applications in the field of Galactic archaeology (Miglio et al. 2017). PLATO can potentially observe a large number of faint red giants. The number of targets that can be observed in addition to the targets of the core program is nevertheless limited to about 40,000 per pointing. An optimal choice of those targets can rely on the seismic performance tool of Mosser et al. (2019). On the other hand, the design and the development of seismic analysis pipelines that are able to process in an automatic way a large number of red giants require the generation of simulated light-curves representative of such stars.

To our knowledge the light-curve simulator developed by De Ridder et al. (2006) in the framework of the Eddington space project is the first code made available to the community that simulates solar-like oscillations together with the stellar granulation noise and the instrumental sources of noise. This simulator relies on a description of the modes and stellar granulation noise that predates CoRoT and *Kepler* space missions. However, our knowledge of solar-like oscillations and stellar granulation has greatly improved since that time. Very recently, Ball et al. (2018) proposed a light-curve simulator dedicated to the TESS mission and that includes an up-to-date description of solar-like oscillators and the granulation background. However, in this simulator, white noise is the only non-stellar source of noise; this means that systematic errors are not included. However, the latter, which are very specific to a given instrument and its space environment, are in general frequency dependent and can only be realistically quantified with simulations made at detector pixel level. Furthermore, the level of the white noise (random noise) also strongly depends on the implemented photometry method and the performance of the instrument. Finally, these simulators do not include planetary transits and are not suited for red giant stars. Indeed, red giants show the presence of numerous mixed-modes, and calculating mixed-mode frequencies with pulsation codes requires a very high number of mesh points in the stellar models thus making the massive generation of corresponding simulated light-curves numerically challenging.

The PLATO mission has some characteristics that make it very different from other space-based mission based on high-precision photometry such as CoRoT, *Kepler* or TESS. Indeed, one of the main specificities of the mission is that it relies on a multi-telescope concept. Among the 26 cameras that compose the instrument, two of them are named 'fast' cameras and work at a 2.5 s cadence while the remaining 24 are named 'normal' cameras and work at a 25 s cadence. The normal cameras are divided into four groups of six cameras, with large fields of view ($\sim$ 1,100 square degrees) that partially overlap. Each camera is composed of four Charge Couple Devices (CCD hereafter) which are read out at the cadence of 25s with a time-shift of 6.25s between each of them. Accordingly, the observations made for a given target by various groups of camera will be time-shifted thereby allowing us to perform super-Nyquist seismic analysis (Chaplin et al. 2014). Because of the large field of view and the long-term change of the pointing direction of each individual camera, star positions will slowly drift on the camera focal plane by up to 1.3 pixels during the 3-month uninterrupted observation sequences. As a consequence, stars will slowly leave the aperture photometry (i.e. masks), leading obviously to a long-term decrease of their measured intensities. Furthermore, during the life of the mission, the instrument will be continuously exposed to radiation (mostly proton impacts). This will generate more and more traps in the CCD thus increasing the Charge Transfer Inefficiency (CTI hereafter, see e.g. Massey et al. 2014, and references therein) over time. Coupled with the long-term drift of the stellar positions, the CTI will induce an additional long-term variability of the photometric measurements.

To mitigate the flux variations induced by the instrument and the observational conditions, the aperture masks used onboard will be updated on a quasi-regular basis. This will nevertheless leave residual flux variations of about several % over three months, which remain high w.r.t. the science requirements. The residual flux variations will fortunately be corrected a posteriori on-ground on the basis of the knowledge of the instrumental point spread function (PSF). Nevertheless, such a correction will leave systematic errors in the power spectrum that will rapidly increase with decreasing frequency. All of these instrumental systematic errors together with the stellar activity noise component can in principle impact the detection and characterisation of the planetary transits, limit the seismic analysis of very evolved red giant stars, and affect any science analysis of the signal at rather low frequencies.

The Plato Stellar Light-curve Simulator [3] (PSLS) aims at simulating stochastically-excited oscillations together with planetary transits, stellar signal (granulation, activity) and instrumental sources of noise that are representative of the PLATO cameras. The simulator allows us to simulate two different types of oscillation spectra: i) oscillation spectra computed on the basis of the so-called Universal Pattern by Mosser et al. (2011) optionally including mixed-modes following the asymptotic gravity mode spacing (Mosser et al. 2012b) and ii) oscillation spectra computed using a given set of theoretical frequencies pre-computed with the ADIPLS pulsation code (Christensen-Dalsgaard 2008).

---

[3] The PSLS source code is available for download from the PSLS website (http://psls.lesia.obspm.fr) as well as from Zenodo.org (http://doi.org/10.5281/zenodo.2581107). The source code is free: you can redistribute it and/or modify it under the terms of the GNU General Public License (for more details see http://www.gnu.org/licenses). The present paper describes the version 0.8.

The instrumental noise level is quantified by carrying out realistic simulations of the instrument at CCD pixel level using the Plato Image Simulator (PIS) for three-month observation sequences. These simulations are performed for different stellar magnitudes, and for both the beginning of life (BOL[4]) and end of life (EOL[5]) observation conditions. The photometry is extracted from these simulated images in the same manner as planned on-board, that is using binary masks that minimise the noise-to-signal ratio (NSR) of each target. The corresponding simulated light-curves are then corrected using PSFs reconstructed on the basis of a microscanning technique, which will be operated in-flight before each three-month observation sequence and which we also simulate in the present work. This set of simulated light-curves, corrected for the instrumental errors, then enables us to quantify the expected level of residual systematic errors. These simulations are then used to derive – as a function of the stellar magnitude – a parametric model of the residual errors in the time domain. This model is in turn implemented into PSLS.

Finally, the other components of the stellar signal (granulation signal, and planetary transits) are included in PSLS following prescriptions found in the literature.

## 2. General principle

The stochastic nature of the different phenomena (i.e. white noise, stellar granulation and stochastically-excited oscillations) are simulated following Anderson et al. (1990, see also Baudin et al. (2007)). As detailed below, the properties of the simulated stellar signal are first modelled in the Fourier domain, we next add a random noise to simulate the stochastic nature of the signal, and finally we perform an inverse Fourier transform to come back into the time domain and derive the corresponding time-series (i.e. light-curve). We note that other authors (e.g. Chaplin et al. 1997; De Ridder et al. 2006) have proposed instead to work directly in the time domain. Although, rigorously equivalent, it is more convenient to describe the stellar signal in the Fourier domain since this is the common way signals (such as pulsation, granulation, and activity) are analysed in solar-like pulsators.

Let $\mathcal{F}(\nu)$ be the Fourier Transform (FT hereafter) of the simulated light-curve $\mathcal{S}(t)$, and $\overline{\mathcal{P}}(\nu)$ the expectation of the Power Spectral Density (PSD) associated with the stellar signal (i.e. the PSD one would have after averaging over an infinite number of realisations). If the frequency bins of the PSD are uncorrelated, we can then show that

$$\mathcal{F}(\nu) = \sqrt{\overline{\mathcal{P}}} \, (u + i \, v) \,, \tag{1}$$

where $u$, and $v$ are two uncorrelated Normal distributions of zero mean and unit variance, and $i$ is the imaginary unit ($i^2 = -1$). We finally compute the inverse Fourier Transform of $\hat{F}(\nu)$ to derive the simulated light-curve $\mathcal{S}(t)$ for a given realisation. We note that the PSD $\mathcal{P}(\nu)$ associated with a given realisation verifies

$$\mathcal{P}(\nu) = |\mathcal{F}(\nu)|^2 = \overline{\mathcal{P}} \left( u^2 + v^2 \right) \,. \tag{2}$$

Our PSD is "single-sided", which means that the integral of the PSD from $\nu = 0$ (excluded) to the Nyquist frequency is equal to the variance of the time-series.

Here, the expectation $\overline{\mathcal{P}}(\nu)$ is the sum of an activity component $\mathcal{A}(\nu)$, the granulation background $G(\nu)$, and the oscillation spectrum $O(\nu)$, that is

$$\overline{\mathcal{P}}(\nu) = \mathcal{A}(\nu) + G(\nu) + O(\nu) \,. \tag{3}$$

In accordance with our initial hypothesis, all these components are uncorrelated. However, some interferences can in principle exist between the various stellar signal components, such as the activity, the granulation and the oscillations. For instance there are some observational evidences about correlations between granulation (i.e. convection) and modes. Indeed, solar mode profiles slightly depart from symmetric Lorentzian profiles (Duvall et al. 1993). Likewise, pieces of evidence for similar asymmetries were recently found in stars observed by *Kepler* (Benomar et al. 2018). Helioseicmic data clearly show that this asymmetry is reversed between velocity and intensity measurements (e.g. Duvall et al. 1993; Nigam et al. 1998; Barban et al. 2004). This reversal is believed to be the signature of a correlation between convection and oscillations (Roxburgh & Vorontsov 1997; Nigam et al. 1998). However, the departures from symmetric Lorentzian profiles are small w.r.t. the mode linewidths. Hence, we consider this as an indication of a small level of correlation between convection (i.e. granulation) and oscillations. Finally, concerning possible interferences between activity and convection, to our knowledge there are no pieces of evidence. For these reasons, in this work, we decided to neglect the correlations between the stellar signal components.

Once the FT associated with the stellar signal is simulated on the basis of Eq. 1, we perform an inverse Fourier transform to come back into the time domain. This then provides the stellar signal as a function of time. However, in order to take into account the fact that each group of cameras are time-shifted by $\Delta t = 6.25$ s, we multiply Eq. 1 by the phase term $e^{i2\pi\Delta t}$ prior to calculating its inverse Fourier Transform.

The instrumental signal component (i.e. the systematic errors plus the instrumental random sources of noise) is simulated in the time domain as explained in Sect. 4. Finally, once the instrumental signal is simulated, it is multiplied by the stellar signal and the planetary transit (which as the instrumental component is simulated in the time domain) to get finally the simulated light-curve averaged over a given number of cameras. We describe in the following sections the way each simulated component is modelled.

## 3. Solar-like oscillations

In this section, we describe the modelling of the oscillation spectrum $O(\nu)$. It is the sum over the different normal modes

$$O(\nu) = \sum_i \mathcal{L}_i(\nu) \,, \tag{4}$$

where each individual resolved mode of frequency $\nu_i$ is described by a Lorentzian profile

$$\mathcal{L}_i(\nu) = \frac{H_i}{1 + \left(2 \left(\nu - \nu_i\right) / \Gamma_i\right)^2} \,, \tag{5}$$

where $H_i$ is the mode height, and $\Gamma_i$ its linewidth. A mode is considered to be resolved when $\Gamma_i > 2\delta f$ where $\delta f$ is the frequency resolution (or equivalently the inverse of the observation duration). In contrast, for an unresolved mode the profile is given by (see, e.g. Berthomieu et al. 2001),

$$\mathcal{L}_i(\nu) = \frac{\pi \Gamma_i H_i}{2\delta\nu} \, \text{sinc}^2 \left[\pi \left(\nu - \nu_i\right)\right] \,, \tag{6}$$

---

where $\delta\nu$ is the resolution of the spectrum.

To go further, one needs to determine the mode frequencies, heights, and line-widths. To do so, we consider two different methods for the frequencies. For main-sequence and subgiant stars, the method consists in computing a set of theoretical mode frequencies using the ADIPLS adiabatic pulsation code while for red giant stars the method developed by Mosser et al. (2011), which relies on what is commonly known as the Universal Pattern. This distinction is motivated by the difficulty to compute red giant frequencies. Indeed, for evolved stars, a proper modelling of the normal frequencies requires an important number of grid points in the innermost layers. While still feasible, this makes the computation more demanding. We therefore adopt a more flexible and affordable method based on asymptotic considerations to ensure the possibility of using the simulator on a massive scale.

### 3.1. Main-sequence and sub-giant stars

The oscillation spectrum is constructed using a set of theoretical eigenfrequencies computed using the ADIPLS code (Christensen-Dalsgaard 2008). The program allows one to include uniform rotational splittings as specified by an input surface rotation period $T_{rot} = 2\pi/\Omega_{surf}$ where $\Omega_{surf}$ is the surface rotation rate. The set of frequencies included in the model are

$$\nu_{n,\ell,m}^{(0)} = \nu_{n,\ell} + \frac{m}{T_{rot}} \left(1 - c_{n,\ell}\right), \qquad (7)$$

where $n$ is the radial order, $\ell$ the angular (or harmonic) degree, $m$ the azimuthal order, and $c_{n,\ell}$ the Ledoux constant (Unno et al. 1989, see, e.g.) provided by ADIPLS. We consider all the modes from $n = 1$ up to the cut-off frequency, with angular degrees ranging from $\ell = 0$ to 3 inclusive. Near-surface effects are eventually added using the empirical correction proposed by Sonoi et al. (2015):

$$\nu_{n,\ell,m} = \nu_{n,\ell,m}^{(0)} + a\,\nu_{max} \left(1 - \frac{1}{1 + \left(\nu_{n,\ell,m}^{(0)}/\nu_{max}\right)^b}\right), \qquad (8)$$

where $a$ and $b$ are two parameters, which are expressed in terms of $T_{eff}$ and $\log g$ thanks to the scaling laws provided in Eqs. 10 and 11 of Sonoi et al. (2015), respectively.

The mode height of each given mode is computed according to

$$H_{n,\ell,m} = G(\nu_{n,\ell,m}; \delta\nu_{env})\, V_\ell^2\, r_{n,\ell,m}^2(i)\, H_{max}, \qquad (9)$$

where $V_\ell$ is the mode visibility ($V_0 = 1$, $V_1 = 1.5$, $V_2 = 0.5$, $V_3 = 0.05$), $H_{max}$ the mode height at the peak frequency, and $r_{n,\ell,m}$ the (relative) visibility of a mode of azimuthal order $m$ within a multiplet for a given inclination angle $i$. The ratio $r_{n,\ell,m}$ is computed according to Dziembowski (1971, see also Gizon & Solanki (2003)) and represents − at fixed values of $n$ and $\ell$ − the ratio of the mode height for a given inclination angle $i$ to the mode height at $i = 0°$. Finally, $G$ is the Gaussian envelope defined by

$$G(\nu_{n,\ell,m}; \delta\nu_{env}) = \exp\left[\frac{-(\nu_{n,\ell,m} - \nu_{max})^2}{\delta\nu_{env}^2/4\ln 2}\right], \qquad (10)$$

where $\delta\nu_{env}$ is the full width at half maximum, which is supposed to scale as (Mosser et al. 2012a):

$$\delta\nu_{env} = 0.66\,\nu_{max}^{0.88}. \qquad (11)$$

This scaling relation was established for red giants. The applications presented in Sect. 6 show that it provides rather good results for less evolved stars.

To compute Eq. 9, we now need to specify $H_{max}$. For a single-side PSD, the mode height is related to the mode linewidth as (see, e.g. Baudin et al. 2005)[6]

$$H_{max} = \frac{2\,A_{max}^2}{\pi\,\Gamma_{max}}, \qquad (12)$$

where $A_{max}$ is the *rms* of the mode amplitude at the peak frequency. The latter is related to the bolometric amplitude $A_{max,bol}$ using the correction proposed for *Kepler*'s spectral band by Ballot et al. (2011)

$$A_{max} = A_{max,bol} \left(\frac{T_{eff}}{5934\,\text{K}}\right)^{-0.8}. \qquad (13)$$

We note that the CoRoT spectral band results in very similar corrections (see Michel et al. 2009). Finally, $A_{max,bol}$ is derived from the scaling relations derived by Corsaro et al. (2013) and defined as

$$\begin{aligned}
\ln(A_{max,bol}) =\ & \ln(A_{max,bol,\odot}) + (2s - 3t)\ln(\nu_{max}/\nu_{max,\odot}) + \\
& (4t - 4s)\ln(\Delta\nu/\Delta\nu_\odot) + \\
& (5s - 1.5t - r + 0.2)\ln(T_{eff}/T_{eff,\odot}) + \ln(\beta) \quad (14)
\end{aligned}$$

where $A_{max,bol,\odot} = 2.53$ ppm (rms) is the maximum of the bolometric solar mode amplitude (Michel et al. 2009), and $s$, $t$, $r$ and $\beta$ are coefficients that depend on the star's evolutionary status (see Tables 3 & 4 in Corsaro et al. (2013)).

Finally, one needs to specify the mode line-widths. To this end, we note that the product of the mode line-width and the mode inertia has a parabolic shape (Belkacem et al. 2011, see Fig. 2). Therefore,

$$\Gamma_{\ell,m} = \Gamma_{max} \left(\frac{I_{max}}{I_{n,\ell}}\right) \gamma(\nu_{n,\ell,m}), \qquad (15)$$

where $I_{n,\ell}$ is the mode inertia, $I_{max}$ is the mode inertia of the radial modes interpolated at $\nu = \nu_{max}$, $\Gamma_{max}$ is the mode linewidth at $\nu = \nu_{max}$ derived from two different scaling relations (see below), and the function $\gamma(\nu)$ models the frequency dependence of the product $\Gamma_{n,\ell,m}I_{n,\ell}$ around $\nu_{max}$. The latter is modelled empirically as follows

$$\gamma(\nu) = 1 + A\,\left(1 - G(\nu_{n,\ell,m}; 2\delta\nu_{env})\right), \qquad (16)$$

where $G$ is the Gaussian function defined by Eq. 10, $A$ is a constant, and $\delta\nu_{env}$ is given by the scaling relation of Eq. 11. With $A = 2$ for $\nu \geq \nu_{max}$ and $A = 6$ for $\nu < \nu_{max}$, Eq. 16 reproduces rather well the variation with frequency of the *solar* mode linewidths. Given the objectives targeted by the simulator, we assume that this empirical description is sufficiently representative for other stars. An alternative approach would have been to use the relation describing the frequency dependence derived from *Kepler* observations by Appourchaux et al. (2014, see its corrigendum in Appourchaux et al. (2016)). However, this relation was established for a limited number of targets and hence in limited ranges in effective temperatures, surface gravities and surface metal abundances. Therefore, to avoid extrapolations we prefer to adopt Eq. 15. In addition, the relation inferred by

---

[6] The additional factor of two comes from the fact we assume here a single-sided PSD while Baudin et al. (2005) assumed a double-sided one.

Appourchaux et al. (2014) was established on limited frequency intervals. Since the mode line-widths scale as the inverse of the mode inertia (Eq. 15), this scaling relation allows us instead to derive the frequency dependence of $\Gamma_{n,\ell,m}$ for the whole acoustic spectrum of a given star.

Finally, the mode line-width at the peak frequency, $\Gamma_{\max}$, is determined on the basis of the scaling relation derived by Appourchaux et al. (2012) from main-sequence *Kepler* targets, that is

$$\Gamma_{\max} = \Gamma_{\max,0} + \beta \left( \frac{T_{\mathrm{eff}}}{T_{\mathrm{eff},\odot}} \right)^s , \qquad (17)$$

where $\Gamma_{\max,0} = 0.20 \, \mu\mathrm{Hz}$, $\beta = 0.97$, and $s = 13.0$.

### 3.2. Red-giant stars

Each mode frequency $\nu_{n,\ell}$ is computed according to the Universal Pattern proposed by Mosser et al. (2011)

$$\nu_{n,\ell,m} = n + \frac{\ell}{2} + \varepsilon(\Delta\nu) - d_{0\ell}(\Delta\nu) + \frac{\alpha_\ell}{2} \left( n - \frac{\nu_{\max}}{\Delta\nu} \right)^2 \Delta\nu + \\ m \, \delta\nu_{\mathrm{rot}} + \delta_{n,\ell} , \qquad (18)$$

where $\varepsilon$ is an offset, $d_{0\ell}$ the small separation, $\alpha_\ell$ the curvature, $\Delta\nu$ the large separation, $\delta\nu_{\mathrm{rot}}$ the rotational mode splitting (included only for dipolar modes, as will be explained later on), and finally $\delta_{n,\ell}$ a term that accounts for a possible coupling with the gravity modes, which results in the deviation of the mode frequency from its uncoupled solution ("pure" acoustic mode) and gives the mode its mixed-mode nature. For a dipole mode, $\delta_{n,\ell}$ is computed according to the asymptotic gravity-mode spacing (Mosser et al. 2012b)

$$\delta_{n,\ell} = \frac{\Delta\nu}{\pi} \arctan \left[ q \tan \pi \left( \frac{1}{\Delta\Pi_1 \nu_{n,\ell}} - \epsilon_g \right) \right] , \qquad (19)$$

where $q$ is the coupling coefficient, $\Delta\Pi_1$ the asymptotic period spacing of the (pure) dipole $g$ modes, and $\epsilon_g$ an offset fixed to the value 0.25, which is representative for most red giants (Mosser et al. 2017). For radial modes, one obviously has $\delta_{n,0} = 0$, while for all modes with angular degree $\ell \geq 2$ we neglect the deviation and assume $\delta_{n,\ell} = 0$.

The mode height of each given mode $(n, \ell, m)$ is given by

$$H_{n,\ell} = G(\nu_{n,\ell}) \, V_\ell^2 \, H_{\max} , \qquad (20)$$

where $G(\nu_{n,\ell})$ is given by Eq. 15, $V_\ell$ is the mode visibility determined from Mosser et al. (2012a) and $H_{\max}$ is the maximum of the mode heights derived from the scaling relation established by de Assis Peralta et al. (2018), that is

$$H_{\max} = 2.01 \, 10^7 \, \nu_{\max}^{-1.9} . \qquad (21)$$

Concerning the mode linewidths $\Gamma_{n,\ell}$, they are assumed to be constant with frequency. This assumption is motivated by the fact that modes are observed in a relatively small frequency range compared to main-sequence and sub-giant stars. This constant value is determined from the theoretical scaling relation of Vrard et al. (2018), which depends on the effective temperature, $T_{\mathrm{eff}}$, and stellar mass as follows

$$\Gamma_{\max} = \Gamma_{\max,0} \left( \frac{T_{\mathrm{eff}}}{4800 \, \mathrm{K}} \right)^{\alpha_T} , \qquad (22)$$

where $\Gamma_{\max,0} = 0.1 \, \mu\mathrm{Hz}$ and $\alpha_T$ is a coefficient which depends on the stellar mass range (see Vrard et al. 2018). The dipolar mixed modes have, however, much smaller line-widths than their associated "pure" acoustic modes. This is mainly because their inertia is much larger as a consequence of the fact they behave as gravity modes in the inner layers. Indeed, the mode line-width scales as the inverse of the mode inertia (see, e.g., Belkacem & Samadi 2013). Let $I_{n,\ell}^m$ (resp. $\Gamma_{n,\ell}^{(m)}$) be the mode inertia (resp. mode line-width) of a dipolar mixed-mode and $I_{n,\ell}^0$ (resp. $\Gamma_{n,\ell}^{(0)}$) that of a "pure" acoustic mode of the same radial order. We then have

$$\Gamma_{n,\ell}^{(m)} = \Gamma_{n,\ell}^{(0)} \left( \frac{I_{n,\ell}^0}{I_{n,\ell}^m} \right) , \qquad (23)$$

where according to our previous assumption $\Gamma_{n,\ell}^{(0)} = \Gamma_{\max}$ for any couple $(n, \ell)$. In Eq. 23, it is assumed that radiative damping in the radiative interior of red giants is negligible. The validity of this assumption has been thoroughly investigated by Grosjean et al. (2014).

To go further, we use the following relation from Goupil et al. (2013):

$$\frac{I_{n,\ell}^0}{I_{n,\ell}^m} \simeq 1 - \frac{I_{\mathrm{core}}}{I} = 1 - \zeta \qquad (24)$$

where $I_{\mathrm{core}}$ is the contribution of the core to the mode inertia, and $\zeta$ is calculated according to Eq. 4 in Gehan et al. (2018). Finally, the rotational splitting for dipolar modes (the term $\delta\nu_{\mathrm{rot}}$ in Eq. 18) is computed on the basis of Eq. 22 in Goupil et al. (2013) by neglecting the surface rotation (see e.g. Mosser et al. 2015; Gehan et al. 2018). Accordingly, we have

$$\delta\nu_{\mathrm{rot}} = \frac{\zeta}{2} \left( \frac{\Omega_{\mathrm{core}}}{2\pi} \right) , \qquad (25)$$

where $\Omega_{\mathrm{core}}$ is the core rotation rate (in rad/s).

The oscillation spectrum is then constructed by summing a Lorentzian profile for each mode. We include modes with radial orders ranging from $n = 1$ up to $n = $ integer $(\nu_c/\Delta\nu)$, where $\nu_c$ is the cutoff-frequency (see Eq. 28), and with angular degrees from $\ell = 0$ to $\ell = 3$.

The simulator requires three main input parameters, $\nu_{\max}$, $T_{\mathrm{eff}}$ and $\Delta\nu$, from which all the other parameters are established using scaling relations, except $\Delta\Pi_1$ and $q$ which can be provided as optional inputs (otherwise no mixed modes are included). In case $\Delta\nu$ is not provided, it is computed according to the scaling relation (Mosser et al. 2013)

$$\Delta\nu = 0.274 \, \nu_{\max}^{0.757} . \qquad (26)$$

The stellar mass used for the granulation scaling relations is determined by combining the scaling relation for $\nu_{\max}$ and $\Delta\nu$ (see Belkacem 2012; Mosser et al. 2010, and references therein):

$$m = M_\odot \left( \frac{\nu_{\max}}{\nu_{\max,\odot}} \right)^3 \left( \frac{\Delta\nu}{\Delta\nu_\odot} \right)^{-4} \left( \frac{T_{\mathrm{eff}}}{T_{\mathrm{eff},\odot}} \right)^{3/2} . \qquad (27)$$

Finally, the cutoff frequency $\nu_c$ is derived from the following scaling relation:

$$\nu_c = \nu_{c,\odot} \frac{g}{g_\odot} \sqrt{\frac{T_{\mathrm{eff},\odot}}{T_{\mathrm{eff}}}} , \qquad (28)$$

where $\nu_{c,\odot} = 5300 \, \mu\mathrm{Hz}$.

## 4. Instrumental errors

Our objective here is to quantify the instrumental sources of error, namely the systematic error and the random noise, and to implement them into PSLS. For the former, a set of simulations at CCD pixel level is carried out while for the random noise we rely on the work made by Marchiori et al. (2019) as explained in Sect. 4.4.

### 4.1. The Plato instrument

PLATO is composed of 24 cameras (named normal cameras) working at a cadence of 25 s and two cameras (named fast cameras) working at a cadence of 2.5 s. Each group of cameras is composed of six normal cameras that see half of the full field of view (2,200 square degrees). The fast cameras point towards the centre of the field of view, and provide the platform with pointing errors for the Attitude Control System. Four Charge Coupled Devices (CCDs) are mounted on the focal plane of each camera. The pixels have a size of 18 $\mu$m and their projected size in the sky represents approximately 15 arcsec.

Every three months, the platform is rotated by 90° in order to maintain the solar panel in the direction of the Sun. Due to the thermal distortion of the platform, changes in the pointing direction of each individual camera are expected during the uninterrupted three-month observation sequences. These variations will lead to long-term star drifts on the focal plane of up to 0.8 pixels in three months. Furthermore, because of the large field of view, the kinematic aberration of light will induce drifts of the stellar positions of up to 0.5 pixels in three months at the edge of the field of view. Both effects add together and result in drifts of up to 1.3 pixels in three months (in the worst case, at the edge of field of view).

### 4.2. The Plato Image Simulator

To quantify the instrumental systematic errors, we generate time-series of small imagettes with the Plato Image Simulator (PIS). This simulator, developed at the LESIA-Observatoire de Paris since the early phases of the PLATO project, has very similar capabilities as the PLATOSim code (Marcos-Arenal et al. 2014). PIS can simulate imagettes representative of PLATO CCDs. It includes various sources of perturbations, such as shot-noise (photon noise), readout noise, background signal, satellite jitter, long-term drift, smearing, digital saturation, pixel response non-uniformity (PRNU), intra pixel response non-uniformity (IPRNU), charge diffusion, and charge transfer inefficiency (CTI). Since our goal is to quantify systematic errors, we turned off all random sources of noise in our instrumental simulations, except in the calculation of the NSR, see Sect. 4.4 ; these are the shot-noise, the readout-noise, and the satellite jitter. CTI is simulated following Short et al. (2013) and activated for end-of-life (EOL) simulations only. Charge diffusion within the CCD pixels is not activated because we still lack a reliable estimate of its amplitude (see the discussion in Sect. 7).

To take into account the impact of long-term drifts of the stellar positions, simulations are generated over 90 days and include a linear drift of 1.3 pixels in three months. To be more realistic, the instrumental point spread functions (PSF) used during these simulations include optical manufacturing errors and integration and alignment tolerances to the nominal design for the nominal focus position. These input PSFs do not include effects due to the detector or the spacecraft (such as the satellite jitter). However, most of them (like PRNU, IPRNU, CTI, and satellite jitter) are in any case included in PIS.

| Parameters | | Value |
|---|---|---|
| Reference flux at V=11 | BOL | 2.17 $10^5$ e-/exp. |
| | EOL | 2.13 $10^5$ e-/exp. |
| Sky background | | 120 e-/s/pixels |
| PRNU | | 1.00% |
| IPRNU | | 0.50% |
| Integration time | | 21s |
| Readout time | | 4s |
| Gain | | 25 e-/ADU |
| Electronic offset | | 1000 ADU |
| Photon noise | | disabled |
| Readout noise e- | | disabled |
| Satellite jitter | | disabled |

**Table 1.** Simulation parameters used with the PIS code.

### 4.3. Simulation parameters and data sets

The flux of each simulated star behaves differently according to their magnitude, position over the CCD, and even position within a pixel (hereafter named intra-pixel position). In order to cover the largest combination of these factors, we use PIS to run 630 artificial star simulations using a combination of:

- 9 stellar magnitudes (from V=9 to V=13 with a step of 0.5),
- 14 focal plane positions over the focal plan (from 1.41° to 18.08° from the optical centre),
- 5 intra-pixel positions for each of the 14 focal plane positions.

The simulations are carried out using the parameters relevant for BOL and EOL conditions. Thus, regarding EOL simulations, the CTI is enabled and the mean optical transmission is assumed to be lower than the BOL one. The CTI model used by PIS requires specifying the number of trap species and their characteristics in terms of density, release time, and cross sections. To this end, Prod'homme et al. (2016) have studied CTI on a representative PLATO CCD that has been irradiated on purpose. This study allowed the authors to identify four trap species and to calibrate their corresponding parameters. We used the parameters derived by Prod'homme et al. (2016). However, the trap densities are rescaled so that the level of CTI reaches the mission specifications at the EOL. The adopted values of the simulation parameters are reported in Table 1.

### 4.4. Photometry extraction

Of the ~ 120,000 targets observed by each camera during a given pointing, about 14,000 of them will have their 6×6 imagettes downloaded on-ground at a cadence of 25 s. For these targets, the photometry will be extracted on-ground on the basis of more sophisticated methods, which are not yet fully established. The photometry of the remaining targets will necessarily have to be performed on-board.

Before computing the photometry, we start with a basic pre-processing of the imagettes aiming to subtract the electronic offset and the background, convert ADU to electrons using the gain, and finally subtract the smearing for each column of the imagette.

Photometry extraction is performed on-board by integrating the stellar flux over a collection of pixels called the *aperture*

or the *mask*. Different strategies for determining the most adequate aperture shape have been the subject of a detailed study (Marchiori et al. 2019), leading to the adoption of binary masks as the best compromise between NSR and stellar contamination ratio. For a given target, its associated binary mask is defined as the subset of the *imagette* pixels giving the minimum noise-to-signal ratio. It is computed through the following scheme

1. Arrange all pixels $n$ from the target *imagette* in increasing order of noise-to-signal ratio $\text{NSR}_n$

$$\text{NSR}_n = \frac{\sqrt{\sigma_{F_{T_n}}^2 + \sum_{k=1}^{N_C} \sigma_{F_{C_{n,k}}}^2 + \sigma_{B_n}^2 + \sigma_{D_n}^2 + \sigma_{Q_n}^2}}{F_{T_n}}. \quad (29)$$

2. Compute the aggregate noise-to-signal $\text{NSR}_{agg}(m)$, as a function of the increasing number of pixels $m = \{1, 2, 3, \ldots, 36\}$, stacking them conforming to the arrangement in the previous step and starting with the pixel owning the smallest $\text{NSR}_n$

$$\text{NSR}_{agg}(m) = \frac{\sqrt{\sum_{n=1}^{m} \left( \sigma_{F_{T_n}}^2 + \sum_{k=1}^{N_C} \sigma_{F_{C_{n,k}}}^2 + \sigma_{B_n}^2 + \sigma_{D_n}^2 + \sigma_{Q_n}^2 \right)}}{\sum_{n=1}^{m} F_{T_n}}. \quad (30)$$

3. Define as the aperture the collection of pixels $m$ providing minimum $\text{NSR}_{agg}(m)$.

In Eq. 29 and Eq. 30, $F_T$ is the target star's mean flux, $\sigma_{F_T}$ the target star's photon noise, $F_C$ the contaminant star's mean flux, $\sigma_{F_C}$ the contaminant star's photon noise, $\sigma_B$ the background noise from the zodiacal light, $\sigma_D$ the overall detector noise (including readout, smearing and dark current noises) and $\sigma_Q$ the quantization noise. Figure 1 illustrates how the NSR typically evolves as the binary mask gets larger following the above scheme. We note that the noise due to satellite jitter is not included in the definition of the mask (Eq. 30). Including the contribution of the jitter noise in the definition of the mask is not trivial because its contribution depends on the final shape of the mask (see e.g. Fialho & Auvergne 2006). Nevertheless, it turns out that for PLATO, the jitter noise is small enough that its does not play a role in the mask shape, the dominant sources of noise being the photon noise for brighter stars and the background and readout noise for fainter stars. Accordingly, once the mask is defined, we include the jitter a posteriori in the estimation of the NSR.

The NSR was estimated for a large sample of targets (so far about 50, 000) with magnitudes ranging from 9 to 13. The targets and their associated contaminant stars were extracted from the Gaia DR2 catalogue (Gaia Collaboration et al. 2018). In total about 3.5 million contaminant stars with magnitudes up to G=21 were included in the calculation.

The calculation of the NSR takes into account the various sources of noise described above and also the fact that the shape of the PSF varies across the field of view. The latest version of the instrument parameters were also considered (details will be given in Marchiori et al. 2019). Typical values of the NSR are given in Table 2 for a single camera and 24 cameras as a function of the PLATO magnitude, P, which is defined in Marchiori et al. (2019) and is by definition directly connected to the flux collected by a PLATO camera. For comparison with the mission specifications (Rauer et al. 2014), we also provide the V magnitude, which is defined here as the flux collected in the Johnson V filter for a reference PLATO target of $T_{\text{eff}} = 6,000$ K.

**Table 2.** NSR as a function of target V and P magnitudes. The values are given for a single camera and for 24 cameras, and were extracted from Marchiori et al. (2019). The rightmost column gives the photon noise limit, that is the NSR one would have if we were limited only by the photon noise of the target.

| V | P | NSR 1 camera [ppm.hr$^{1/2}$] | NSR 24 cameras [ppm.hr$^{1/2}$] | Photon noise limit 24 cameras [ppm.hr$^{1/2}$] |
|---|---|---|---|---|
| 8.1 | 7.76 | 51.9 | **10.6** | 10.5 |
| 8.5 | 8.16 | 63.2 | **12.9** | 12.7 |
| 9.0 | 8.66 | 80.3 | **16.4** | 16.0 |
| 9.5 | 9.16 | 101.9 | **20.8** | 20.1 |
| 10.0 | 9.66 | 130.8 | **26.7** | 25.4 |
| 10.5 | 10.16 | 169.0 | **34.5** | 32.2 |
| 11.0 | 10.66 | 219.5 | **44.8** | 40.8 |
| 11.5 | 11.16 | 290.0 | **59.2** | 52.0 |
| 12.0 | 11.66 | 387.5 | **79.1** | 66.1 |
| 12.5 | 12.16 | 523.2 | **106.8** | 84.3 |
| 12.9 | 12.56 | 678.5 | **138.5** | 102.6 |

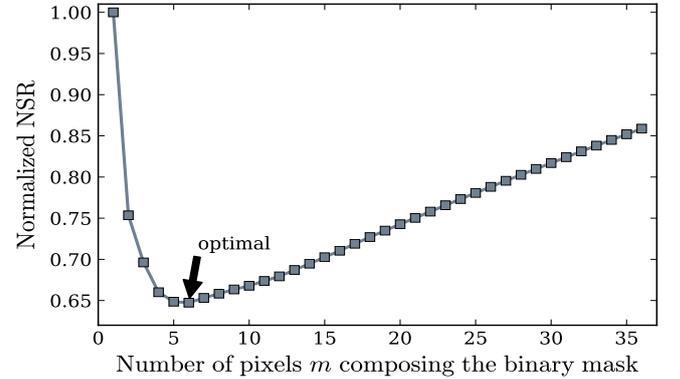

**Fig. 1.** Typical NSR evolution curve as a function of increasing aperture size. Pixels with the smallest noise-to-signal ratio are added-up successively. The collection of pixels giving the lowest aggregate NSR defines the binary mask for a given target.

For comparison, we also reported the values of the NSR in the photon noise limit (i.e. when there is only the photon noise due to the target). The relative contribution of random noises that add (quadratically) to the target photon noise increase with increasing stellar magnitude from 20 % at magnitude V=8.5 up to 65 % at V=13.

Unless the NSR value is imposed by the user, the latter is obtained by interpolating the values given in Table 2 for a given V magnitude.

### 4.5. Mask update

An example of a light-curve obtained with a fixed optimal binary mask is shown in Fig. 2 (top) for a target of magnitude

V=11. Because of the long-term drift of the star and the fact that the mask is maintained at the same position during the three-month observation sequence, we observe a significant long-term decrease of the stellar flux. In this worst case scenario (a displacement of 1.3 pixels in three months), the flux decreases by about 15 %, which subsequently results in an increase of the NSR by about 8 % (see Fig. 2 – bottom). The NSR increase obviously has an impact on the science objectives of the mission, in particular on the planet detection rate. Indeed, a higher NSR at a given magnitude reduces the number of targets for which the NSR is lower than a given threshold, which subsequently lowers the number of detected planets.

To mitigate the impact of the long-term drift and to maintain the NSR as low as possible, the proposed solution is to update the mask when required during the three-month observation sequences. An example is shown in Fig. 2 (top). We see in this example that the peak-to-peak variation of the flux is maintained within 4 % (top panel) while the variation of the NSR is maintained within less than 2 % (lower panel). Therefore, the mask updates always guarantee that one reaches the best possible NSR. Furthermore, it is also found that the mask updates partially mitigate the impact of CTI.

It is interesting to note that some mask updates simultaneously reduce the flux and the NSR. This is because the NSR does not scale linearly with the flux. Indeed, due to the presence of the readout noise and the complex shape of the PSF, two masks collecting the same amount of flux can have a different number of pixels and hence different contributions of the readout noise.

Obviously, each mask update will introduce a discontinuity at a well known instant, which for a given target will be different from one camera to another. For instance in the example shown in Fig. 2, the binary mask has been updated seven times. It is however possible to reduce the number of updates by increasing the threshold above which a variation of the NSR since the last update must trigger a mask update. Furthermore, as explained below, the discontinuities induced by each update as well as the long-term flux variations induced by the long-term drifts can be efficiently corrected a posteriori on-ground.

### 4.6. Light-curve correction

Knowing the PSF at any position across the field of view and the time displacements of a given target, it is possible – given its associated aperture mask – to reconstruct a synthetic light-curve, which exactly mimics the time variation of the star flux induced only by the long-term drift over the CCD plane as well as the discontinuity induced by each mask update. The light-curve correction then consists in building such a synthetic light-curve assuming that the star has an unit intensity, then dividing the real stellar light-curve by the synthetic one. The quality of this correction intrinsically depends on our ability to construct the PSF and to derive the stellar displacement in time. As explained in detail in Sect. 4.7, the stellar PSFs will be reconstructed during the in-flight calibration phases on the basis of a microscanning sequence coupled with a dedicated inversion method.

Concerning the stellar displacements, the fast camera will provide information about the short-term variations of the satellite attitude (i.e. satellite jitter) with a high cadence (2.5 s) and a sufficient accuracy. This information will then be directly translated in terms of the short-term variations of the attitude of each given normal camera. Concerning the long-term displacements, centroids of a larger set of targets will be measured using the imagettes registered on-board at a cadence of 25 s. These centroids will be used to derive the attitude of each camera at any instant. The combination of the two sets of information will finally provide both the short-term (i.e. jitter) and long-term time-displacements of any target.

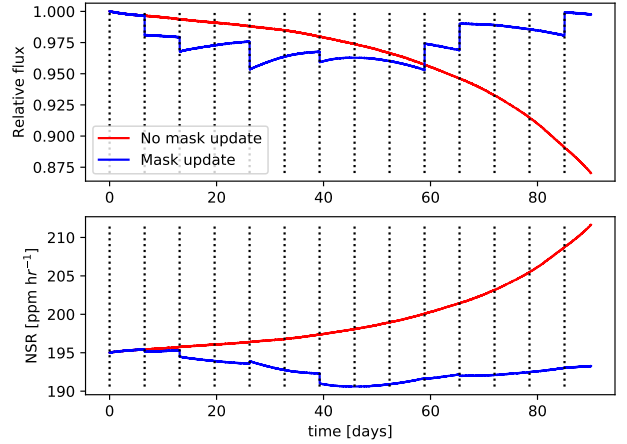

**Fig. 2. Top:** Examples of light-curves generated from a simulated time-series of CCD imagettes. The red curve corresponds to the light-curve generated with a fixed binary mask while the blue one includes a series of mask updates. The dotted vertical lines identify the times at which the masks were updated. **Bottom:** Corresponding time variations of the NSR.

Finally, it has been established that the PRNU is a limiting factor for this correction. However, prior to the launch, the PRNU will be measured with an accuracy better than 0.1 % (rms), which is sufficient to leave a negligible level of residual error in the corrected light-curves (Samadi 2015).

Two examples of corrected (individual) light-curves are displayed in Fig. 3 for a V=11 PLATO target. The upper light-curve corresponds to BOL observation conditions while the lower one to the EOL. It is clearly seen that the residual flux variations are larger at the EOL: for that particular target the peak-to-peak flux variations is as high as about 1 % at the EOL while it remains within about 0.2 % at the BOL. This is explained by the combined effect of the CTI and the star drift. Indeed, as the star moves, the energy distribution in the different pixels vary with time and so does the CTI. This effect is named differential CTI.

The small discontinuities seen in the light-curve occur each times the mask has been updated. These discontinuities are of the order 500 ppm. Hence, they remain small compared to the photon noise, which is about 2,000 ppm for this target and for a single camera. It is also worth noting that for a given target observed with several cameras the instants at which the mask updates occur are different between the different cameras. Accordingly, the systematic errors induced by these updates are uncorrelated between the cameras and accordingly, their impacts on the final light-curves (obtained after averaging several individual light-curves) will be significantly reduced (see Sect. 4.10).

### 4.7. Point spread function reconstruction

One of the challenges with the PLATO mission is the relatively large size of the camera pixels (approximately 15 arcsec as projected on the field-of-view) compared to the typical size of the point spread functions (PSFs).[7] Accordingly, raw camera pictures do not provide a sufficient resolution of the PSFs, thus requiring the use of a specific strategy in order to obtain the PSFs with a sub-pixel resolution. In the PLATO mission, the adopted strategy is similar to the one applied to *Kepler* observations (Bryson et al. 2010): a microscanning sessions in which a series of imagettes with sub-pixel displacements are obtained (Green 2011). High resolution PSFs will then be reconstructed by inverting the imagettes along with a precise knowledge of the displacements. Such PSFs will be obtained for a number of reference stars across the field-of-view. The PSF at any position will then be obtained via interpolation using the reference PSFs. The resultant PSFs will subsequently be used in correcting the light-curves sent down by PLATO as explained in Sect. 4.6.

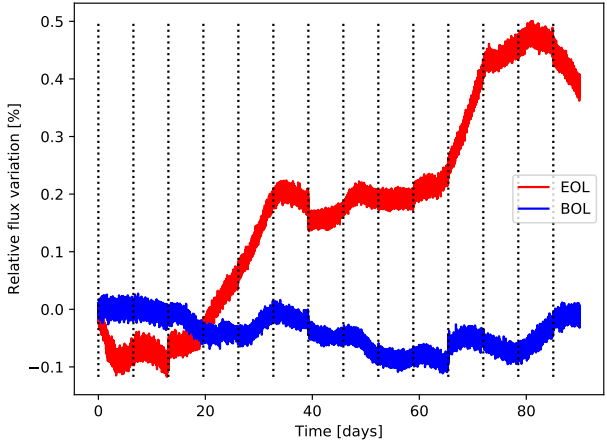

**Fig. 3.** Examples of single-camera light-curves generated from a simulated time-series of CCD imagettes and after correcting it for the long-term drift of the stellar position as explained in Sect. 4.6. We represent the relative flux variation in %. The red (reps. blue) curve corresponds to EOL (resp. BOL) observation conditions. The dotted vertical lines identify the times at which the masks were updated.

### 4.7.1. Microscanning sessions

The microscanning sessions will typically last for 3 hours and lead to a series of 430 imagettes composed of 6 × 6 pixels encompassing the target stars. The telescope will be pointing in a slightly different direction for each imagette resulting in small sub-pixel displacements of the target stars. A continuous microscanning strategy has been opted for, that is the position will be changing continuously throughout the manoeuvre rather than stopping for each imagette and then starting again (Ouazzani et al. 2015). The displacements do not need to fulfil stringent criteria in order to be suitable for the inversions, but only to form a path which roughly covers a pixel uniformly (for more details see Reese 2018a). Accordingly, this path has primarily been determined based on technological constraints. However, a precise knowledge of the displacements is essential for carrying out successful inversions. Various tests have shown that the fast cameras

---

[7] We note that the point spread function changes significantly across the relatively large field-of-view (about 20° in radius).

are able to obtain this information from the centre-of-brightness of reference stars.

The displacements will form an Archimedean spiral such that the distance, $D$, between consecutive images is approximately constant, and the distance between consecutive spiral arms is $D\sqrt{3}/2$, thus leading to the formation of near-equilateral triangles depending on the relative positions of imagettes on consecutive arms. Furthermore, the spiral needs to approximately cover 1 pixel. The combination of these constraints leads to a spiral like the one illustrated in Fig. 4.

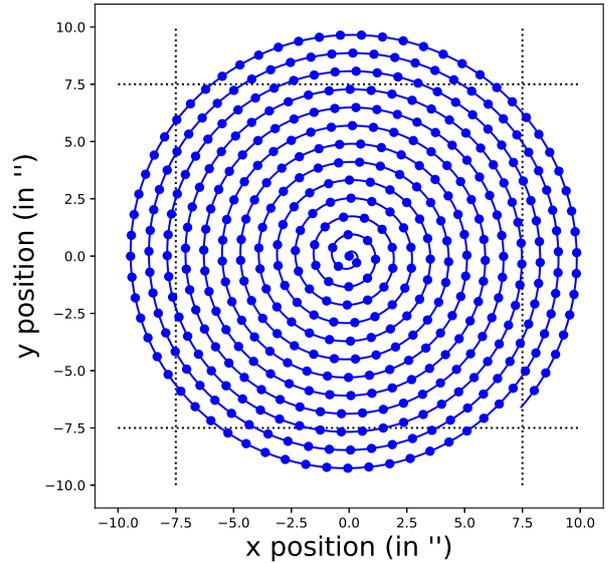

**Fig. 4.** Archimedean spiral used for the microscanning strategy. The dotted lines correspond to the CCD pixel borders.

### 4.7.2. PSF inversions

In order to carry out the inversions, it is necessary to discretise the PSF by expressing it as a sum of basis functions:

$$f(x, y) = \sum_i a_i \phi_i(x, y), \qquad (31)$$

where the $a_i$ are unknown coefficients which will be determined via the inversion, and the $\phi_i$ basis functions. Typical choices of basis functions include sub-pixel indicator functions, or Cartesian products of cubic B-splines. The typical resolution used for these basis functions is $1/20^{\text{th}}$ of a pixel (along both the $x$ and $y$ directions), given the number of imagettes from the microscanning session. This high resolution representation of the PSF then needs to be integrated over the pixels of the imagettes. Equating the resultant integrals with the observed intensities in these pixels leads to the following equation

$$Ax = b, \qquad (32)$$

where $b$ is a vector composed of the observed intensities from the imagettes, $x$ a vector composed of the coefficients $a_i$, and $A$ the discretisation matrix. The inverse problem is then to extract the high resolution PSF, $x$, knowing $A$ and $b$. Given that the number of unknowns does not necessarily equal the number of observables, this problem needs to be inverted in a least-squares sense. Furthermore, some form of regularisation needs to be included in order to obtain well-behaved solutions. Finally, the resultant high-resolution PSF needs to remain positive. A sufficient (though not necessary) condition for this is to impose that the coefficients $a_i$ are positive, provided the basis functions $\phi_i$ remain positive.

Two inversion techniques have been used for solving Eq. 32. The first is an iterative approach called the Multiplicative Algebraic Reconstruction Technique (MART, Censor 1981; Green & Wyatt 2006), which starts from a positive smooth solution and iteratively corrects it using one constraint at a time. Given that the corrections are applied in a multiplicative manner, the solution remains positive. The number of iterations is then used to control the degree of regularisation. The second approach is a regularised least-squares approach with a positivity constraint on the coefficients. The regularisation term consists of a 2D Laplacian multiplied by a weight function which leads to a higher amount of regularisation in the wings of the PSF. Accordingly, cubic B-splines are used with this approach given that these are continuously twice-differentiable. This term is then multiplied by a tunable regularisation parameter. As shown in Reese (2018b), this second approach leads to better results in most cases (see an example in Fig. 5 bottom panel) and is accordingly the preferred approach.

### 4.8. Analysis of systematic errors in terms of PSD

We compute the PSD associated with each corrected light-curve, both for the BOL and EOL data sets. Two examples are shown in Fig. 6 at the BOL and the EOL for a star located at a given position in the field of view and at the sub-pixel position $P_0$ (pixel corner). These PSDs are compared with the PLATO requirements in terms of allowed systematic errors at V=11. At the EOL, the requirements are marginally exceeded in the frequency range [10 $\mu$Hz - 100 $\mu$Hz]. This is a consequence of the presence of CTI. We stress that we expect to be able to correct for the CTI. However, this correction is not yet fully modelled and hence cannot yet be reliably quantified. Accordingly, the predictions for the EOL have to be considered conservative at this moment.

We find that the PSD of the residual light-curve can satisfactory be fitted with a function of the form:

$$\mathcal{I}(\nu) = H_1 \left(\frac{\nu_1}{\nu}\right)^{\alpha_1} + \frac{H_2}{1 + (2\pi\tau_2\nu)^{\alpha_2}}, \qquad (33)$$

where $H_1$, $\alpha_1$, $H_2$, $\tau_2$ and $\alpha_2$ are the fitted parameters, and $\nu_1 = 1/T_1$ with $T_1 = 90$ days (three months). For convenience, we further define $\sigma^2$ as the variance of the residual light-curve, which is also the integral $\mathcal{I}(\nu)$. The quantity $\sigma$ corresponds to the amplitude of the systematic errors and is related to the other parameters according to the relation

$$\sigma^2 = \frac{H_1\nu_1}{\alpha_1 - 1} + \frac{H_2}{2\tau_2 \sin(\pi/\alpha_2)}. \qquad (34)$$

We fit each residual light-curve with the function given by Eq. 33. Depending on the parameters, we find that the parameter values significantly vary with the sub-pixel positions. This is not surprising since about 90 % of the star's intensity is concentrated in a square of 2.2×2.2 pixels (an example of such a PSF is displayed in Fig. 5). As a result, a change of the sub-pixel position of the star's centroid induces important changes in the charge distribution. In contrast, at fixed sub-pixel positions, changes of the parameter values with the stellar field of view are in general weaker. Finally, the parameters controlling the amplitudes of the systematic errors (namely $H_1$ and $H_2$) are found to strongly vary with the stellar magnitude.

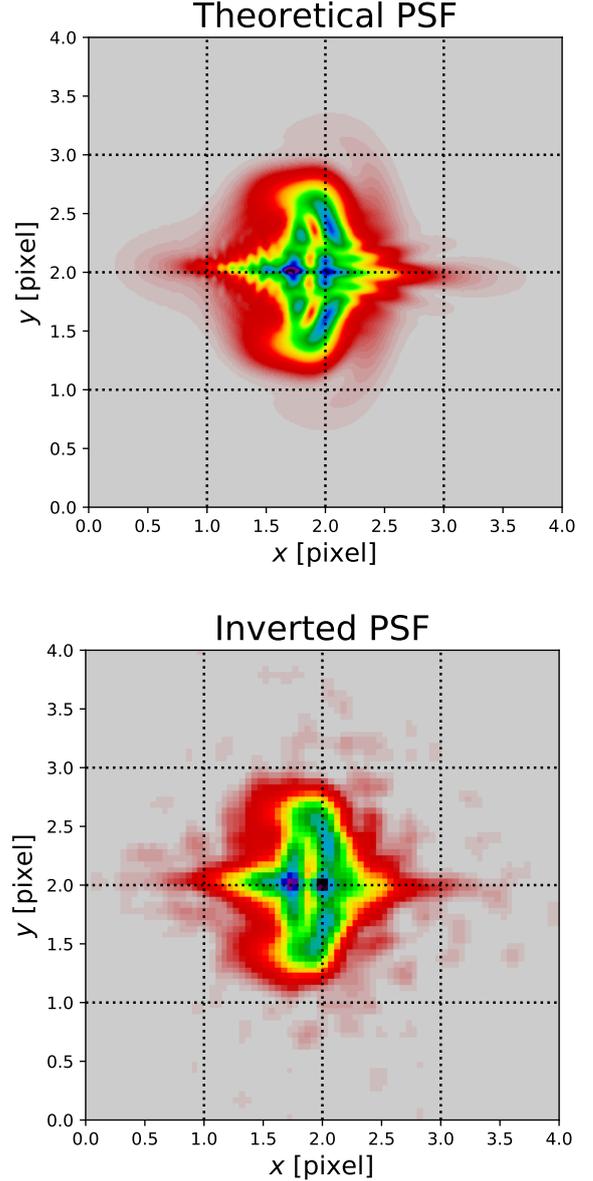

**Fig. 5. Top:** Original PSF. **Bottom:** PSF obtained by inversion on the basis of the microscanning technique (see Sect. 4.7). The dotted lines correspond to the CCD pixel borders.

Fig. 7 highlights the impact of the star magnitude. Indeed, the 90th percentile of the quantity $\sigma$ [in ppm] is displayed as a function of the star magnitude, both for BOL and EOL conditions. In general, the residual systematic errors increase with increasing star magnitude. As expected, the systematic errors at EOL are systematically much higher than at BOL by about a factor ten at magnitude 8.5 and down to about a factor four at magnitude 13. However, they hardly exceed 0.5 % (rms) in the magnitude range considered here.

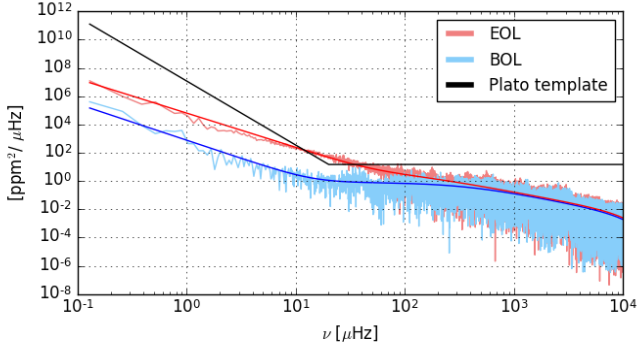

**Fig. 6.** PSD of the residual light-curve obtained with two three-month PIS simulations, one representative of the BOL (cyan) and the second of the EOL (pink). The results shown here correspond to a star of magnitude 11. The solid coloured line represents the fitted model defined in Eq. 33. The solid black line represents the PLATO requirements in terms of systematic errors translated for a single camera by assuming that they are uncorrelated between the different camera (see text).

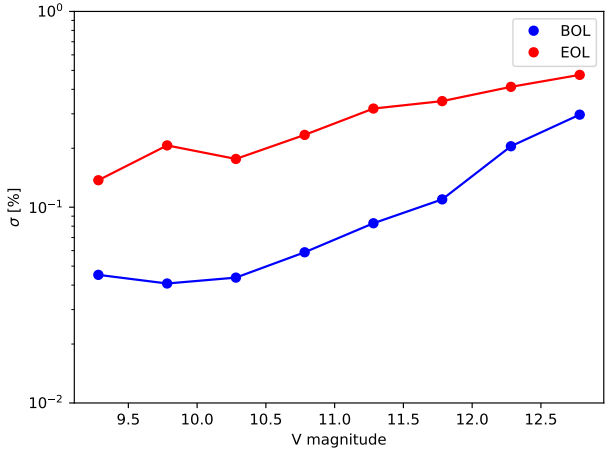

**Fig. 7.** Amplitude of the systematic errors ($\sigma$) as a function of the stellar magnitude. This quantity is computed according to Eq. 34. At each stellar magnitude, only the 90th percentile is displayed.

### 4.9. Modelling the systematic errors in the time domain

The systematic errors were analysed in the previous section in terms of PSD because this is the most convenient way to compare directly with the mission requirements in terms of allowed systematic errors. Indeed, the latter are specified in terms of the PSD. However, modelling the systematic errors this way has the obvious consequence of destroying the phase of the instrumental or operational perturbations (e.g. the discontinuities induced by the mask updates or other effects). While this is in principle not a problem when the stellar signal is analysed in terms of the PSD (e.g. as this is typically done for the granulation or the solar-like oscillations), this can be misleading for the analysis taking place in the time domain, as for instance the detection and the characterization of planetary transits. To overcome this, we decide instead to model the systematic errors in the time domain.

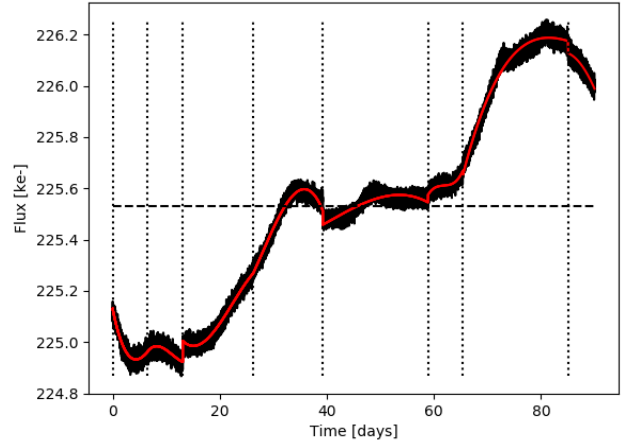

**Fig. 8.** Example of a generated instrumental light-curve (single camera) fitted with the piecewise polynomial decomposition of Eq. 35. The black line represents the generated light-curve and the red one the result of the fit.

Due to the quasi-regular mask updates, the residual light-curve is piecewise continuous (each piece corresponding to an interval of time where the aperture mask is unchanged). We find that each piece can be well reproduced by a third order polynomial. Accordingly, we decompose each generated instrument light-curve as follows

$$s(t) = \bar{s} \sum_{i=1}^{N} \Pi\left(\frac{t - t_i}{d_i}\right)\left(1 + p_{3,i} + p_{2,i}x + p_{1,i}x^2 + p_{0,i}x^3\right), \quad (35)$$

where $\bar{s}$ is the light-curve time-average, $N$ the number of masks used for a given imagette time-series, $i$ the mask index, $t_i$ the time the mask is first applied or updated, $d_i$ the time during which it is maintained, $x \equiv (t - t_i)/\tau_0$, $\tau_0$ a time constant (set arbitrarily to 90 days), $p_{j,i}$ the polynomial coefficients associated with the mask $i$, and finally $\Pi(x)$ a function defined as

$$\Pi(x) = \begin{cases} 1 & \text{if } 0 \leq x < 1 \\ 0 & \text{if } x < 0 \text{ or } x \geq 1. \end{cases}$$

While the coefficient $p_{3,i}$ informs us about the amplitude of the discontinuity induced by a given mask $i$, the three other coefficients ($p_{0,i}$, $p_{1,i}$ and $p_{2,i}$) inform us about the long-term variations of the instrument residuals obtained with that mask.

Each of the generated instrument light-curve is fitted by the model given by Eq. 35. An example of such a fit is given in Fig. 8. In most cases, this polynomial model reproduces very well the main characteristics of the systematic errors, in particular the jumps induced by the mask updates as well as the long-term variations induced by the long-term star drifts.

### 4.10. Implementation into PSLS

The model for the systematic errors presented in the previous section is implemented into PSLS as follows: we have at our disposal a set of $p$ coefficients for each stellar magnitude, focal plane position (i.e. PSF), and sub-pixel position. We first identify the positions (focal plane and sub-pixel) corresponding to the magnitude the closest to that of the star we want to simulate. For each position, the number of $p$ coefficients depends on the number of masks used at that position. Then, each individual light-curve simulated by PSLS is divided into quarters. For each quarter, we randomly select the set of coefficients $p$ among the ensemble previously selected. We proceed in the same way for each quarter and for each individual camera. By proceeding this way, we simulate the fact that each star will have different PSFs and sub-pixel positions in the different cameras and that these PSFs and sub-pixel positions will change after the rotations of the spacecraft by 90° every 3 months. An example of such simulated light-curves is shown in Fig. 9.

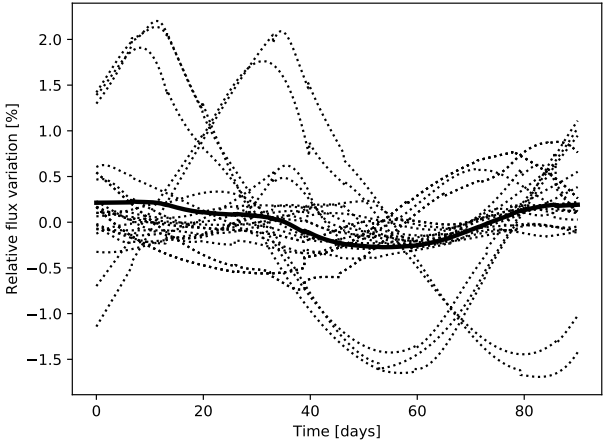

**Fig. 9.** Simulated instrument residual light-curves (systematic errors) over 90 days for a star of mangitude V=11 and for the EOL conditions. The light-curves are plotted in terms of relative variations and were generated using Eq. 35 and as explained in Sect. 4.10. Each dotted line corresponds to an individual light-curve (here 24 in total) while the thick solid line corresponds to the light-curve obtained by averaging the 24 simulated light-curves.

The above example also illustrates the benefit of averaging the light-curves over several cameras. Indeed as shown in Fig. 9 averaging over, for example 24 cameras, substantially reduces the residual errors because the systematic errors are not always in phase and the mask updates do not always occur at exactly the same times. However, the figure also highlights some degree of correlation between the individual light-curves. Indeed, it can clearly be seen that some light-curves are close to being in phase. These correlations are expected as explained and discussed in Sect. 7.2.

## 5. Other signal components

### 5.1. Stellar granulation

The granulation background is simulated by assuming two pseudo-Lorentzian functions

$$G(\nu) = \sum_{i=1,2} \frac{h_i}{1 + (2\pi\tau_i\nu)^{\beta_i}}, \qquad (36)$$

where $h_i$ is the height, $\tau_i$ the characteristic time-scale, and $\beta_i$ the slope of the Lorentzian function. The values of $h_i$ and $\tau_i$ are determined from the scaling relations established by Kallinger et al. (2014) with *Kepler* observations of red giants, sub-giants and main-sequence stars. These scaling relations are a function of peak frequency $\nu_{max}$ of the oscillations and the stellar mass $M$. Following Kallinger et al. (2014), the values of the two slopes ($\beta_1$ and $\beta_2$) are both fixed to four.

### 5.2. Stellar activity

The stellar activity signal is simulated assuming a Lorentzian function

$$\mathcal{A}(\nu) = \frac{2\,\sigma_A^2\,\tau_A}{1 + (2\pi\tau_A\nu)^2}, \qquad (37)$$

where $\sigma_A$ is the amplitude and $\tau_A$ is the characteristic time-scale of the activity component. Both parameters have to be specified by the user (but see the discussion in Sect. 7).

### 5.3. Planetary transit

Planetary transit light-curve are simulated on the basis of Mandel & Agol (2002)'s formulation and using the Python implementation by Ian Crossfield at UCLA[8]. This model allows us to specify several parameters controlling the characteristics of the transit light-curve. Among them, PSLS allows us to specify the planet radius, the orbital period, the semi-major axis and finally the orbital angle. We have adopted a quadratic limb-darkening law (cf. Section 4 of Mandel & Agol (2002)) and assumed default values for the corresponding two coefficients (namely $\gamma_1 = 0.25$ and $\gamma_2 = 0.75$). However, these coefficients can be set by the user.

## 6. Simulated stellar oscillation spectra

As a preliminary remark, we stress that the goal of the simulator is not to provide state-of-the-art modelling of a given target but rather to be able to mimic the main characteristics of objects of interest. Accordingly, we did not carry out a quantitative and extensive comparison between outputs of our simulator and light-curves (or equivalently PSD) obtained from high-precision photometric observations, from space missions such as CoRoT and *Kepler*. However, to illustrate the quality and the relevance of the simulated light-curves, we performed a qualitative comparison with *Kepler* observations. Three *Kepler* targets were selected according to the quality of the data and their evolutionary status: a main sequence star (16 Cyg B – KIC 12069449), a sub giant (KIC 12508433) and a giant on the Red Giant Branch (KIC 009882316). For each of them, a simulation was generated with stellar parameters and models as close as possible to the corresponding target.

### 6.1. Main sequence star

16 Cyg B (KIC 12069449) belongs to the *Kepler* legacy sample (Lund et al. 2017a,b). As input for PSLS, we considered a set of theoretical adiabatic mode frequencies computed with ADIPLS, using one of the stellar models considered in Silva Aguirre et al. (2017). The effective temperature and surface gravity were adjusted in accordance with the 1D stellar model while the seismic indices $\nu_{max}$ and $\Delta\nu$ were taken from Lund et al. (2017a).

We generated an initial light-curve assuming a V=10.0 PLATO target observed with 24 cameras in EOL conditions and for a duration of 2 years. The choice of magnitude is motivated by the fact that we expect to derive stellar ages with the required accuracy (10 %) with 24 cameras up to the magnitude

---

[8] http://www.astro.ucla.edu/~ianc/

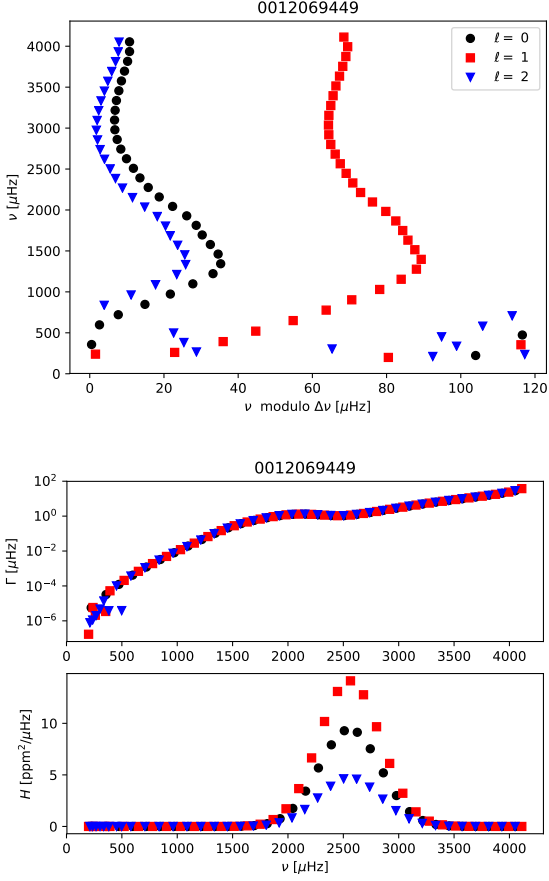

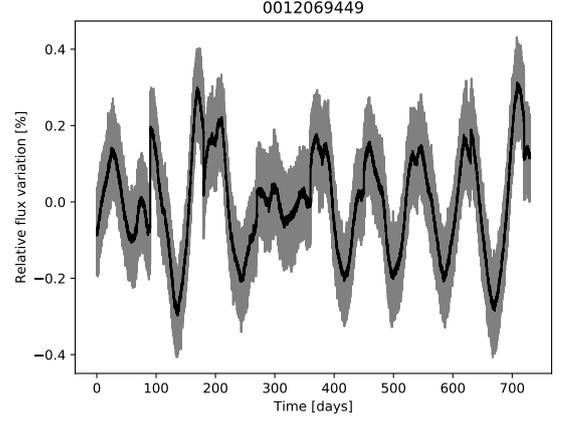

**Fig. 11.** Simulated light-curve for the main sequence star 16 Cyg B (KIC 12069449) with simulation parameters representative of a V=10.0 PLATO target observed in EOL conditions with 24 cameras (see the details given in Sect. 6.1). The grey curve corresponds to the raw light-curve while the black ones is the light-curve averaged over one hour.

**Fig. 10. Top:** Échelle diagram corresponding to the the frequencies used as input for the simulations made for 16 Cyg B (KIC 12069449). A mean large separation of $\Delta\nu = 118.9$ $\mu$Hz was assumed when plotting the échelle diagram. **Bottom:** Corresponding mode linewidths (top) and mode heights (bottom).

V=10.0 (Goupil 2017).[9] The two parameters controlling the activity component have been adjusted so that it matches that of the activity component seen in the 16 Cyg B Kepler light-curve. The corresponding PSLS configuration file is given in Appendix A. The mode frequencies, line-widths, and heights used as input for the simulation are displayed in Fig. 10. The modes for which the frequencies significantly depart from the general trends are mixed modes. However, they have such low amplitudes that in practice they are not at all detectable.

The corresponding simulated light-curve is displayed in Fig. 11 while the corresponding PSD is plotted in Fig. 12, where we have depicted the various contributions to the PSD. As can been seen, the systematic errors start to dominate over the stellar signal only below $\sim 20\,\mu$Hz. On the other hand, they remain negligible in the frequency domain where the solar-like oscillations and stellar granulation signal lie. As expected at that magnitude, the random noise (white noise) dominates the signal in this domain. Nevertheless, the presence of the oscillations in the PSD is clearly discerned when zooming and smoothing the PSD in this frequency domain (see bottom panel of Fig. 12).

The simulated PSD cannot be directly compared with *Kepler* observations for that star because PLATO and *Kepler* have different characteristics and furthermore 16 Cyg B is so bright that its image on the CCD is saturated. Therefore, to perform a comparison we adjusted the white noise level (equivalently the NSR value) so that it matches the level of the white noise seen at high frequency in the *Kepler* light-curve. We compare in Fig. 13 the simulated PSD with the *Kepler* observations. Qualitatively, we note a fair agreement between the simulation and the observations. The figure however highlights some differences, in particular in terms of mode heights and the width of the oscillations envelope. As the characteristics of the oscillations are obtained through scaling relations, we do not except the match to be perfect, and in any case this is not the ultimate goal of the simulator.

### 6.2. Sub-giant star

The sub-giant star KIC 12508433 observed by *Kepler* is among the sub-giant stars studied in detail by Deheuvels et al. (2014). As an input for PSLS, we use the same stellar parameters as in this study as well as the set of theoretical mode frequencies that best fits the seismic constraints. As for 16 Cyg B (KIC 12069449), we adjusted the white noise level so that it matches the level of the white noise seen at high frequency in the corresponding *Kepler* light-curve. The comparison between the simulated PSD and the one computed from the *Kepler* light-curve is shown in Fig. 14. Here also, we have a good qualitative agreement between both PSDs. Nonetheless, some mismatch is visible by eye, especially concerning the mode heights and the width of the oscillation envelope.

### 6.3. Red-giant star

The red giant KIC 9882316 has been studied extensively since the *Kepler* era. Precise measurements of its seismic indexes ($\Delta\nu$, $\nu_{\max}$, $\Delta\Pi$ and $q$) have been published for example in Mosser et al. (2015). We generated for this red giant a simulated light-curve on the basis of the UP method (see Sect. 3.2). The latter requires specifying the seismic indexes as well as the effective tempera-ture ; all these parameters are taken from Mosser et al. (2015). To illustrate the quality of the light-curve expected for red giants with PLATO, we first perform a simulation for a V=12.5 PLATO target observed with 24 cameras in EOL conditions for a duration of 2 years. The corresponding PSLS configuration file is given in Appendix A. We note that solar-like oscillations are also expected to be detectable in fainter red giants, but we limit ourselves to this magnitude because the systematic errors were not quantified for fainter stars. The PSD of the simulated light-curve is displayed in Fig. 15, where we have also plotted the different contributions to the signal. As can been seen, the systematic errors remain negligible compared to the solar-like oscillations and stellar granulation. On the other hand, they dominate below $\nu \sim 20\mu$Hz.

---

[9] This threshold is obviously lower for stars observed with less than 24 cameras.

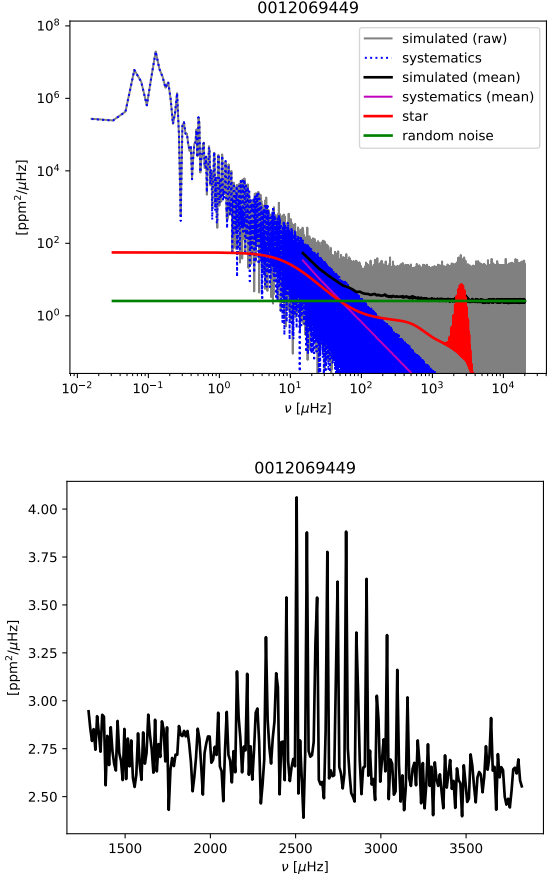

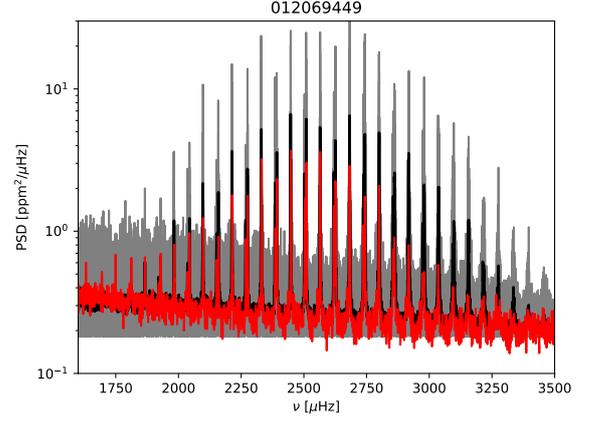

**Fig. 13.** Comparison of a simulated PSD and the one obtained from *Kepler* observations of the main sequence star 16 Cyg B ( KIC 12069449). The grey and black lines have the same meaning as in Fig. 12. The red curve corresponds to the smoothed PSD obtained from the observations.

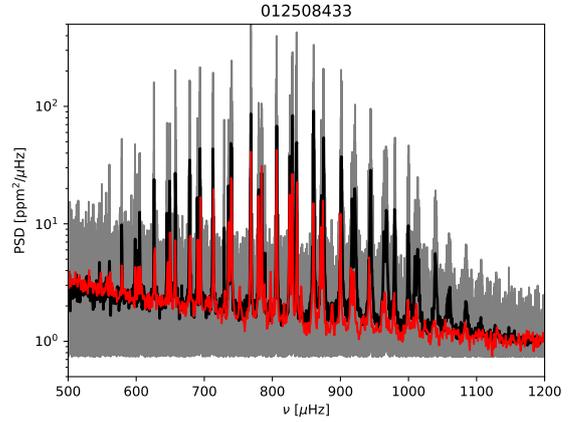

**Fig. 14.** Same as Fig. 13 for the sub-giant star KIC 12508433 observed with *Kepler*.

**Fig. 12.** PSD of the simulated light-curve of 16 Cyg B (KIC 12069449) shown in Fig. 11. **Top:** the full PSD. The grey curve represents the raw PSD (i.e. un-smoothed PSD) while the black line corresponds to the PSD obtained after applying a running average over a width of $10\,\mu$Hz. The coloured lines represent the various contributions to the signal (see the associated legend). **Bottom:** zoom in the frequency domain where solar-like oscillations are detected. Only the smoothed PSD is shown.

Finally, we compare the predictions made by PSLS with *Kepler* observations. We again adjust the white noise level to match the *Kepler* observations for that target and considered a simulation duration of 4 years. The comparison is shown in Fig. 16. The agreement between the simulation and the *Kepler* observations is rather good. In particular, we see that the mixed-mode frequencies and heights are quite well reproduced.

# 7. Discussion

We discuss in this section the limitations of the current approach and possible future improvements.

## 7.1. Instrument model

As far as the modelling of the instrument is concerned, there is still an important effect missing in the image simulator (PIS), which is the Brighter Fatter Effect (BFE hereafter). Indeed, there are several pieces of evidence showing that spot images using CCDs do not exactly scale with the spot intensity: bright spots tend to be broader than faint ones, using the same illumination pattern (see Guyonnet et al. 2015, and references therein). The BFE is fundamentally due to the self-electrostatic interaction between charges in different pixels. This broadening, which mainly affects bright targets, would not be a problem as long as these interactions are stable in time. However this cannot be the case since the long-term drift of the stellar position changes the charge distribution in the different pixels. Analytical mod-els (e.g.. Guyonnet et al. 2015) can be easily implemented into PIS, and that can subsequently be used to correct both the CCD imagettes and the light-curves generated on-board. Such analytical models involve several free coefficients that can be calibrated on-ground with the test bench dedicated to the calibration of the flight PLATO CCDs. The BFE is not expected to evolve with time so that the parameters obtained with the on-ground calibration can be used throughout the mission. The calibration procedure for the PLATO CCD is not yet established but can in principle follow the one proposed in Guyonnet et al. (2015). As soon as we have at our disposal calibrated values of the BFE coefficients, it will be possible to update our simulations and derive new prescriptions to account for this additional source of systematic errors.

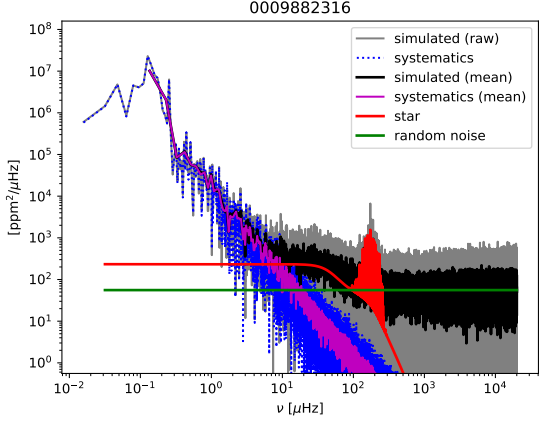

**Fig. 15.** PSD of the simulated light-curve for the Kepler red giant KIC 9882316 seen as a PLATO target of V=12.5 in EOL conditions with 24 cameras. The curves have the same meaning as the top panel of Fig. 12.

Charge diffusion within the CCD was neglected in this work since we still lack reliable estimates of its amplitude in the case of the PLATO CCD. However, charge diffusion is expected to a have non-negligible impact on the performance since it enlarges somewhat the width of the PSF and leads to the suppression of the small-scale structures of the optical PSF. It has been shown for example by Lauer (1999) that in standard rear-illuminated CCDs this phenomena can be modelled by performing the convolution of the optical PSF with a 2D Gaussian kernel with a given width, which strongly depends on the wavelength and type of CCD device. To what extent this model is applicable to the PLATO CCD and what typical width to use for representative PLATO targets are still open questions. Fortunately, tests on representative PLATO CCD are currently taking place at the ESTEC and will provide feedback on this issue that should allow us to improve the present performance assessment.

Besides the above mentioned effects taking place at the detector level, PLATO will be subject to many others perturbations that are not yet taken into account in the present simulator. Among them, we can in particular mention thermal trends after the rotations of the spacecraft by 90° every 3 months, the regular thermal perturbations induced by the daily downlinks, the momentum wheels de-saturations, and finally the residual outliers that would not have been detected by the outlier detection algorithms. All these perturbations are not yet well characterized but will be better known in the future, in particular with the deeper involvement of the prime contractor of the platform in the project and progress in the definition of the data processing pipeline.

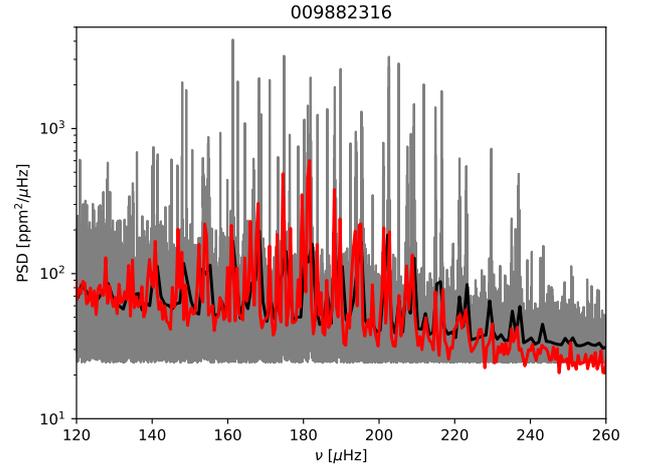

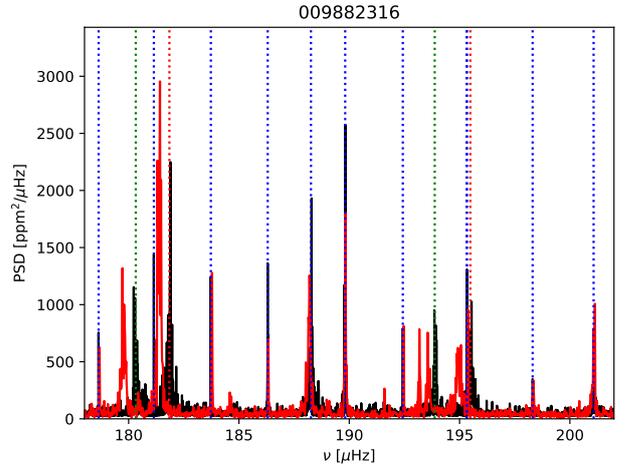

**Fig. 16.** Same as Fig. 13 for the red giant star KIC 9882316 observed with *Kepler*. **Top:** the entire oscillation spectrum. **Bottom:** Zoom around the maximum peak frequency. Here we did not apply a running average over the PSD. The dotted red/blue/green vertical lines represent the frequency locations of the radial/dipole/quadrupole modes respectively.

### 7.2. Single to multiple instrument simulations

Strong correlations between the light-curves coming from different cameras are expected. For instance, stellar drifts along the focal plane are expected to be strongly correlated between the cameras. Although for each given target, the associated PSF and aperture mask can differ between cameras, variations in the stellar flux induced by stellar drifts will present some degree of correlation, which only pixel-level simulations made for *several* cameras can quantify.

PSLS generates the instrument systematic errors of each individual light-curve individually. However, it uses the model parameters derived from pixel-level simulations made for a single camera only (see Sect. 4.9). Each observed target will have different PSFs and sub-pixel positions in the different cameras with which it is observed. Accordingly, to simulate this diversity, PSLS randomly selects the model parameters derived at various positions (for a given star magnitude).

However, this approach is to some extent conservative. Indeed we use the systematic errors evaluated for the same camera. Each individual light-curve will be corrected a posteriori on-ground on the basis of auxiliary data (such as the PSF) obtained by calibrating independently each individual camera. There are good reasons to believe that the systematic errors will be different from one camera to another. Indeed, the systematic errors made on each individual calibration are expected to be different because the cameras are not exactly identical. Indeed, the cameras do not have the exact same alignments of the CCD over the focal plane, same focal plane flatness, same PRNU, same optical manufacturing and alignment errors, etc. However, to confirm this it is required to simulate a statistically sufficient number of cameras with slightly different setups (results from a limited attempt can be found in Deru et al. 2017).

### 7.3. Stellar contamination

While the presence of contaminant stars was taken into account in our calculation of the NSR as a function of the stellar magnitude (see Sect. 4.4), this is not the case for the systematic errors. It is, in principle, possible to take into account the contaminant stars in the reconstruction of the stellar PSF and the generation of the three-month imagette time-series (Reese & Marchiori 2018). However, this is numerically challenging since due to the high diversity in terms of configuration, statistically reliable quantification of the impact of stellar contamination would require a much larger sample of simulations. Accordingly, we plan to perform, in the near future, simulations on the basis of a sufficiently large stellar field extracted from the Gaia DR2 catalogue.

### 7.4. Stellar activity and rotational modulations

Although the presence of stellar activity has little impact on solar-like oscillations, its presence is critical for the detection of planetary transits. At the present time, the parameters of the activity component still need to be specified by the user. Hence, our objective is to implement into the simulator some empirical descriptions of the magnetic activity sufficiently realistic to be representative of solar-like pulsators in the context of PLATO. To this end, we plan to analyse a large set of *Kepler* targets and derive from their spectra, in a similar way as for example in de Assis Peralta et al. (2018), two main characteristic parameters of the activity, namely the characteristic time-scale and the amplitude associated with the activity component. Once these parameters are derived for a large sample of stars, we believe it will be possible to derive some relations between these parameters and some stellar parameters, such as the surface rotation period and the Rossby number which is the ratio of the rotation period and convective turnover time. Indeed, the differential rotation existing at the interface between the convective envelope and the internal radiative zone is believed to be at the origin of the stellar dynamo while convection is believed to be responsible for the diffusion of the magnetic field in the convective zone (see e.g. Montesinos et al. 2001, and references therein).

Finally, one other missing activity-related signal is the rotational modulation due to the presence of rather large spots on the stellar surface. It is hence planned to implement in the near future some of the existing spot models (for a review on this problem see Lanza 2016). However, one difficulty is to have at our disposal representative prescriptions for the model parameters, for instance typically the number of spots, their sizes and their lifetimes. To our knowledge, such prescriptions do not exist yet. Therefore, as a starting point we plan to let the user chose these parameters.

## 8. Conclusion

We have presented here a light-curve simulator, named the PLATO Solar-like Light-curve Simulator (PSLS), that aims at simulating, as realistically as possible, solar-like oscillations together with other stellar signals (granulation, activity, planetary transits) representative of stars showing such pulsations. One of the specificities of this tool is its ability to account for instrumental and observational sources of errors that are representative of ESA's PLATO mission. The latter were modelled on the basis of the Plato Image Simulator (PIS), which simulates the signal at the CCD pixel level. At the Beginning Of Life, we show that the systematic errors are always compliant with the specifications, whereas at the End Of Life they marginally exceed the specifications between $10\,\mu$Hz and $100\,\mu$Hz approximately (see Fig. 6) as a result of Charge Transfer Inefficiency (CTI). However, some mitigation options for the CTI are currently under study (e.g. charge injection, increasing the camera shielding). Although the procedure is not yet fully established, existing correction algorithms can be implemented in the context of PLATO (e.g. Short et al. 2013; Massey et al. 2014).

The PIS code is however not adapted to generating in a massive way simulated long-duration light-curves (e.g. up to two years in the case of PLATO). This is why a parametric description of the systematic errors expected in the time domain has been derived from the PIS simulations. This model reproduces both the residual long-term flux variations due to the instrument as well as the jumps induced by the mask-updates for those of the targets (the large majority of the targets of sample P5) for which photometry is extracted on-board. Implemented into PSLS, this parametric model enables us to mimic in a realistic and efficient way the instrument systematic errors representative of the PLATO multi-telescope concept. Hence, with the inclusion of stellar signal components that are the most representative for the PLATO targets together with a realistic description of the instrument response function, this light-curve simulator becomes an indispensable tool for the preparation of the mission. Its adaptation to other future space missions is in principle possible, provided that some analytical prescriptions for the instrumental and environmental sources of errors representative of the mission are available.

Light-curves simulated with PSLS allow us to conclude that the systematic errors remain negligible above about $100\,\mu$Hz and only start to dominate over the stellar signal below $\sim 20\mu$Hz. Accordingly, they should not impact the core science objectives of PLATO. One the other hand, they can potentially impact the analysis of the signal below $\nu \sim 20\mu$Hz. In both cases, however, firm conclusions deserve dedicated studies, which are beyond the scope of the present work. It must further be made clear that the level of systematic errors predicted by the present modelling is, strictly speaking, only representative for those targets for which the photometry is extracted on-board (i.e. the large majority of the sample P5). For all the other samples, in particular the main sample (P1), the photometry will be extracted on-ground and thus will not suffer from the quasi-regular mask-updates. Therefore, a lower level of systematic errors are expected for these samples. Accordingly, the use of PSLS must be considered as a conservative approach for these samples.

This simulator is based on our current knowledge of the instrument and of the current development of the correction pipeline. Although already well advanced, this knowledge will improve in the near future as soon as a first flight model of the camera will be available and fully characterized (around the beginning of 2021). At that time, it will be relatively easy to update our pixel-level simulations and subsequently the parameters used by the model for the systematic errors as well as the Noise-to-Signal Ratio (NSR) table.

*Acknowledgements.* This work has benefited from financial support by CNES in the framework of its contribution to the PLATO mission. It has made use of data from the European Space Agency (ESA) mission *Gaia* (https://www.cosmos.esa.int/gaia), processed by the *Gaia* Data Processing and Analysis Consortium (DPAC, https://www.cosmos.esa.int/web/gaia/dpac/consortium). Funding for the DPAC has been provided by national institutions, in particular the institutions participating in the *Gaia* Multilateral Agreement. We acknowledge T. Prod'homme and D. Tiphéne for the useful discussions we had about various CCD effects. We thank the anonymous referee for his/her remarks which enabled us to improve significantly the simulator and the present article.

## Appendix A: Configuration file for the main-sequence star KIC 12069449 (16 Cyg B)

```
# PLATO Solar-like Light-curve Simulator (PSLS) configuration file (V 0.8)

# Observations conditions
Observation:
  Duration : 730.  #  [days]
  MasterSeed : 1704040900 # Master seed of the pseudo-random number generator

# Instrument parameters
Instrument:
  Sampling : 25. # Sampling cadence of each camera [s]
  IntegrationTime : 21. # Integration time [s]
  NGroup : 4 # Number of camera groups (1 -> 4)
  NCamera : 6 # Number of camera per group (1->6)
  TimeShift : 6.25 # Time shift between camera groups [s]
  RandomNoise:
    Enable: 1
    Type: PLATO #  either 'User' or 'PLATO'. In the first case the NSR value is specified by the user (see be
    NSR :  73. # Noise to signal ratio [ppm/hr], for one single camera, this value takes into account all
  Systematics:
    Enable : 1
    Table : PLATO_systematics_EOL.npy # The binary file containing  the systematics parameters

# Stellar parameters
Star:
  Mag : 10. #  Magnitude
  ID : 12069449  #  star ID
  ModelDir : # Directory containing the single models or the grid (parameters and associated theoretical f
  ModelType: single   # Type of model: 'grid' or 'single' or 'UP'
  ModelName: 0012069449  # Name of the input model, to be specified when ModelType = 'single', , the progra
  ES : ms  # Evolutionary status: 'ms' for the main-sequence phase, 'sg' for the sub-giant phase, 'rg' for re
  Teff : 5750.  # Effective temperature [K]
  Logg : 4.353 # Surface gravity, ignored for the UP
  numax : 179.3 # frequency of the maximum power [muHZ], used only with the UP
  delta_nu : 13.68 # Mean large separation [muHz], used only for UP, -1 if you want this parameter to be c
  DPI : 80.58 # Asymptotic values of the gravity mode period spacing [s], used only with the UP, -1 if you
  q : 0.15 # Mixed modes coupling factor, used only with the UP
  SurfaceEffects: 1 # Include mode near-surface effects, not implemented for the UP
  SurfaceRotationPeriod :   0. # Surface rotation period [days], not used with the UP
  CoreRotationFreq :   0. # Core rotation frequency [muHz], this is by definition Omega/2pi*1e6 where Omega
  Inclination : 0. # Inclination angle [deg.]

Activity :
  Enable: 1
  Sigma : 40. # Amplitude of the activity component [ppm]
  Tau : 0.2 # Time-scale of the  activity component [days]

Granulation :
  Enable: 1

# Transit parameters
Transit :
  Enable: 0
  PlanetRadius : 0.5   # in jupiter radius
  OrbitalPeriod : 10.  # in days
  PlanetSemiMajorAxis : 1.  # in U.A.
  OrbitalAngle : 0. # in deg
  LimbDarkeningCoefficients: [0.25,0.75]
```

**Appendix B: Configuration file for the red giant star KIC 9882316**

```
# PLATO Solar-like Light-curve Simulator (PSLS) configuration file (V 0.8)

# Observation conditions
Observation:
  Duration : 730. #  [days]
  MasterSeed : 1704040900 # Master seed of the pseudo-random number generator

# Instrument parameters
Instrument:
  Sampling : 25. # Sampling period of each camera [s]
  IntegrationTime : 22. # Integration time [s]
  NGroup : 4 # Number of camera groups (1 -> 4)
  NCamera : 6 # Number of cameras per group (1 -> 6)
  TimeShift : 6.25 # Time shift between camera groups [s]
  RandomNoise:
    Enable: 1
    Type: PLATO #  either 'User' or 'PLATO'. In the first case the NSR value is specified by the user (see be
    NSR :  1. # Noise to signal ratio [ppm/hr], for a single camera. This value takes into account all ran
  Systematics:
    Enable : 1
    Table: PLATO_systematics_EOL.npy # The binary file containing  the systematics parameters

# Stellar parameters
Star:
  Mag : 12.5 #  Magnitude
  ID : 9882316 #  star ID (integer)
  ModelDir : # Directory containing the single models or the grid (parameters and associated theoretical f
  ModelType: UP # Type of model: 'grid' or 'single' or 'UP'
  ModelName: KIC9882316  # Name of the input model, to be specified when ModelType = 'single'
  ES : rg # Evolutionary status: 'ms' for the main-sequence phase, 'sg' for the sub-giant phase, 'rg' for re
  Teff : 5400.  # Effective temperature [K]
  Logg : 3.934 # Surface gravity, ignored for the UP
  numax :  181.77120898 # frequency at maximum power [muHZ], used only with the UP
  delta_nu : 13.70885191 # Mean large separation [muHz], used only with the UP, -1 if you want this parame
  DPI : 80.30739925 # Asymptotic values of the gravity mode period spacing [s], used only with the UP, -1
  q :  0.15611347  # Mixed mode coupling factor, used only with the UP
  SurfaceEffects: 0 # Include near-surface effects in mode frequencies, not implemented for the UP
  SurfaceRotationPeriod :  0. # Surface rotation period [days], not used with the UP
  CoreRotationFreq :  0. # Core rotation frequency [muHz], this is by definition Omega/2pi*1e6 where Omega
  Inclination : 0. # Inclination angle [deg.]

Activity :
  Enable: 0
  Sigma : 1000. # Amplitude of the activity component [ppm]
  Tau : 30. # Time-scale of the activity component [days]

Granulation :
  Enable: 1

# Transit parameters
Transit :
  Enable: 0
  PlanetRadius : 0.5   # in jupiter radii
  OrbitalPeriod : 10.   # in days
  PlanetSemiMajorAxis : 1.   # in A.U.
  OrbitalAngle : 0. # in deg
  LimbDarkeningCoefficients: [0.25,0.75]
```